\documentclass[aps,pra,twocolumn,groupedaddress,showpacs]{revtex4-1}
\usepackage{amsmath, amsthm, amssymb}
\usepackage{enumitem}
\usepackage{color,calc,graphicx,subfig}
\usepackage{svgcolor}
\usepackage{tikz}
\usepackage{rotating}
\usepackage{version}
\usepackage{placeins}

\newcommand{\D}{\displaystyle}

\newtheorem{theorem}{Theorem}

\begin{document}

\title{Practical sharing of quantum secrets over untrusted channels}

\author{Anne Marin}

\author{Damian Markham}

\affiliation{CNRS LTCI, D\'{e}partement Informatique et R\'{e}seaux, Telecom ParisTech, 23 avenue d'Italie, CS 51327,  75214 Paris CEDEX 13, France}


\begin{abstract}

In this work we address the issue of sharing a quantum secret over untrusted channels between the dealer and players.
Existing methods require entanglement over a number of systems which scales with the security parameter, quickly becoming impractical.
We present protocols (interactive and a non-interactive) where single copy encodings are sufficient. Our protocols work for all quantum secret sharing schemes and access structures, and are implementable with current experimental set ups. For a single authorised player, our protocols act as quantum authentication protocols.


\end{abstract}

\maketitle

\section{Introduction}

In secret sharing a dealer wishes to distribute a secret to a network of players such that only authorised sets of players can access the secret, and unauthorised sets of players cannot. After the initial protocols for sharing classical secrets \cite{Shamir79,Blakley79}, ones for sharing quantum secrets were later developed \cite{HBB99,CGL99}, and have found uses including secure multiparty computation \cite{BCG06}. However, these protocols rely on trusted channels between the dealer and the players. In practice, channels may be corrupted either by unavoidable noise, or malicious attacks.

One way to resolve this situation would be to use the quantum authentication protocol \cite{BCGST02}. However this protocol is highly impractical in that it uses error correcting codes and which would require encoding each qubit sent from the dealer into a highly entangled state (or perform entangling measurements, which would allow the generation of large entangled states), the size of which scale with the security parameter. This difficulty, on a par with the coherences needed for quantum computing, renders this approach infeasible in the near future.

In this work we present a protocol which is universal (it works for all quantum secret sharing protocols and access structures) and is implementable with current experimental setups, for example by using graph states. This is possible because our protocol uses only single copy encodings. As in the authentication scheme \cite{BCGST02}, our protocol uses an initial shared secret key between the dealer and receivers. We begin by introducing an interactive protocol, which will serve as a basis for the non-interactive protocol which follows. We then give an example of an explicit graph state implementation for sharing a secret between five players such that any three can access the secret and fewer cannot. We finish with a discussion on possible variants of the protocol including the possibility of abort, and the merits of graph state implementations \cite{MS08,KFMS10,MM13,MMP13}.


\section{Protocols}\label{intro}

In quantum secret sharing, a secret $|\psi\rangle=\alpha|0\rangle + \beta |1\rangle$ is encoded by the dealer $d$ into some logical basis $|\psi_L\rangle_P =\alpha|0_L\rangle_P + \beta |1_L\rangle_P$ on $|P|$ systems, and distributed to the players $P$. A set of players $B\subset P$ are \emph{authorised} if they can access the secret. This is equivalent to there existing a pair of logical operators $X_L$, $Z_L$ which are nontrivial only over the systems $B$, and act over the logical basis in the appropriate way $X_L|i_L\rangle = |i\oplus 1 _L\rangle$, $Z_L|i_L\rangle = (-1)^i |i  _L\rangle$  (where $\oplus$ symbolises sum modulo two) \cite{MM13}. These are used by $B$ to access the secret. A set of players $B\subset P$ are \emph{unauthorised} if they can get no information at all about the quantum secret state. The choice of the logical basis determines the authorised and unauthorised sets. All our protocols are built on the existence of these schemes (which exist for all access structures \cite{CGL99}), and we will use this notation in our protocols.

For our first protocol, we use a general entangled based picture of secret sharing \cite{MM13}.
In this picture the dealer and the players share an entangled EPR state
\begin{eqnarray} \label{EQN: EPR}
|\Phi\rangle_{dP}=\frac{|0\rangle_d|0_L\rangle_P+|1\rangle_d|1_L\rangle_P}{\sqrt{2}},
\end{eqnarray}
which is then used to teleport the secret to the players. The interactive protocol presented below essentially verifies that the dealer and a give set of authorised players $B$ share this state (or the associated reduced state), in which case the teleportation will be successful. More explicitly, the dealer generates many ($S$) copies of the entangled EPR state (\ref{EQN: EPR}) and uses all but one to test the state (steps 3. and 4 in the protocol below), and one to teleport (step 5.). By randomly choosing when to test and when to use the state for teleportation any malicious actions cannot help but be detected. We will see after that this can be translated to a non-interactive protocol by replacing communication by a shared random key.

\bigskip

\noindent{\bf Interactive Protocol}
\begin{enumerate}[topsep=2pt, partopsep=0pt,itemsep=0mm]
\item Dealer $d$ generates $S$ EPR states, $|\Phi\rangle_{dP}^{\otimes S}$, and sends the shares of each one to $P$. \label{PInt: IntAuthProc step1}

\item After $P$ received all their parts, $d$ chooses $r\in [1,...S]$ at random and sends $r$ to $P$.

\item For EPR pairs  $i\neq r$, $d$ chooses $t_i\in [0,1]$ and measures $X_{d}$ if $t_i=0$ or measures $Z_{d}$ if $t_i=1$, and denotes the result $y_i$, and sends $t_i$ and $y_i$ to $P$. \label{PInt: Testa}

\item For EPR pairs $i\neq r$, accessing set $B$ measure $X_{L,B}$ if $t_i=0$ or measures $Z_{L,B}$ if $t_i=1$. Denoting result by $y_i'$ if $y_i=y_i'$ ACCEPT, if $y_i\neq y_i'$ REJECT.\label{PInt: Test}

\item For $i=r$, $d$ uses EPR pair $r$ to teleport the secret state onto the logical basis, denoting the bell basis measurements $x$, and sends $x$ to all $P$. Upon receiving $x$, $B$ decodes using $X_{L,B}$ and $Z_{L,B}$.\label{PInt: Teleportation}
\end{enumerate}

\bigskip

This protocol is effectively a quantum authentication protocol from the dealer to authorised set $B$. In \cite{BCGST02} a framework for quantum authentication is laid out, along with definitions of completeness, soundness and security, which we will adopt here. A general quantum authentication scheme for sending messages from $A$ to $B$ is described by a randomly chosen classical key $k\in \mathcal{K}$ that is shared by $A$ and $B$, and associated encoding operations $A_k$ and $B_k$ respectively. At the end of the protocol $B$ has a system which encodes the message, and a classical register which encodes the decision whether to accept or reject in orthogonal states $|ACC\rangle$ and $|REJ\rangle$.
A quantum authentication scheme is \emph{$\epsilon$-secure} if for all states $|\psi\rangle$ it satisfies the two conditions.
\begin{itemize}
\item {\bf Completeness}. For all keys $k\in \mathcal{K}$
\begin{equation}
B_k(A_k(|\psi\rangle \langle \psi|)) = |\psi\rangle\langle \psi| \otimes |ACC\rangle\langle ACC|.
\end{equation}
\item {\bf Soundness}. For all (possibly malicious) channels $O$, describing the expected state on Bob's side after the protocol as $\rho_B = \frac{1}{|\mathcal{K}|}\sum_{k\in \mathcal{K}}B_k(O(A_k(|\psi\rangle \langle \psi|)))$, and denoting the two projections $P_{fail}^{|\psi\rangle} := (I-|\psi\rangle\langle\psi|) \otimes |ACC\rangle\langle ACC|$, then
\begin{equation} \label{EQN: Security Prob Fail}
Tr \left(P_{fail}^{|\psi\rangle} \rho_B \right)\leq \epsilon.
\end{equation}
\end{itemize}
We say that our protocols are $\epsilon$-secure if all authorised sets $B$ can authenticate the secret with $\epsilon$-security, and unauthorised sets of players get no information. The latter is guarenteed by the use of the original secret sharing logical operators in our protocol, as is the completeness, we will prove the soundness now.

The left hand side of equation (\ref{EQN: Security Prob Fail}) is equal to the probability of accepting multiplied by the fidelity to $B$'s resulting state (averaged over keys) to the space orthogonal to the ideal state $|\psi\rangle$. That is, it represents a failing in the protocol, so we want to make it arbitrarily small (with some security parameter $S$). In order to prove soundness, we will bound this by considering statements about the entangled states themselves, before the teleportation.

We first introduce the operator $\Pi_{dB} := 1/4(I_d \otimes I_B  + X_d \otimes X_{L,B} + Z_d X_d \otimes Z_{L,B} X_{L,B} + Z_d \otimes Z_{L,B} )$, which is a projector onto a space where all states are maximally entangled between $d$ and $B$. More specifically every state in this subspace can be expressed in the form
$\frac{|0\rangle_d|0_{L(i)}\rangle_B + |1\rangle_d |1_{L(i)}\rangle_B}{\sqrt{2}}$, where $\{|0_{L(i)}\rangle_B,|1_{L(i)}\rangle_B\}$ are some basis of $B$ such that $i$ represents a possible logical bases, $X_{L,B}|j_{L(i)}\rangle_B=|{j\oplus1}_{L(i)}\rangle_B$, $Z_{L,B}|j_{L(i)}\rangle_B= (-1)^j |j_{L(i)}\rangle_B$ and we will use many such bases in (\ref{EQN: logical expansion}) such that $\langle j_{L(i)}|k_{L(m)}\rangle_B=\delta_{j,k}\delta_{i,m}$.

Consider the state $\rho_{dB}^r$ which is used to teleport in protocol step \ref{PInt: Teleportation} (conditioned on accepting on all other pairs, see (\ref{EQN: Conditional state})). Any such state can be purified to $|\Psi\rangle_{dBE}$, which can be expanded
\begin{eqnarray}\label{EQN: logical expansion}
|\Psi\rangle_{dBE}&=&(\Pi_{dB}\otimes I_E + (I_{dB}-\Pi_{AB})\otimes I_E))|\Psi\rangle_{dBE}\nonumber\\
&=&\sqrt{F}\left(\sum_i \alpha_i (|0\rangle_d|0_{L(i)}\rangle_B + |1\rangle_d |1_{L(i)}\rangle_B) |\psi_i\rangle_E\right) \nonumber \\
&&+ \sqrt{1-F}|\xi\rangle_{dBE},
\end{eqnarray}
where $F=Tr(\Pi_{dB} \rho_{dB}^r)$. If this state is then used to teleport a state $|\psi\rangle$ from $d$ to $B$, followed by $B$ doing a logical decoding to ancilla system $B'$ (for example $B$ performs a logical swap onto $B'$) the state recovered $\rho_{B'}$ has fidelity $f:= \langle \psi|\rho_{B'}|\psi \rangle$ with the original state satisfying $f\geq F$. Furthermore,
\begin{align}
Tr \left(P_{fail}^{|\psi\rangle} \rho_B \right) &= Tr \left(((I-|\psi\rangle\langle\psi|) \otimes |ACC\rangle\langle ACC|) \rho_B \right) \nonumber \\
&\leq Tr \left(\Pi_{dB}^\perp \otimes |ACC\rangle\langle ACC| \rho_{dB_{AR}} \right),
\end{align}
where $\Pi_{dB}^\perp$ is the projector onto the space orthogonal to $\Pi_{dB}$ and  $\rho_{dB_{AR}} = 1/S \sum_{r=1}^S p_{ACC}^r \rho_{dB}^r\otimes |ACC\rangle \langle ACC | +p_{REJ}^r \rho_{dB}^{r,REJ} \otimes |REJ \rangle \langle REJ|$, $ p_{ACC}^r$ and $p_{REJ}^r$ are the probability of accepting and rejecting respectively when using $r$, and $\rho_{dB}^{r,REJ}$ is the state conditioned on rejecting.

Denoting the POVM element associated to accepting the test step \ref{PInt: Test} for pair $i$ as $M_{ACC_i}=1/2\left(\frac{I_{d_i} \otimes I_{B_i} + X_{d_i} \otimes X_{L,B_i}}{2} + \frac{I_{d_i}\otimes I_{B_i} + Z_{d_i}\otimes Z_{L,B_i}}{2}\right)$, we have $M_{ACC_i}\leq \ \frac{\left(I_{d_i}\otimes I_{B_i} + \Pi_{d_i,B_i}\right)}{2}$. Then, if we call the total state shared over all copies of the dealer and the players $B$, $\rho_{d_1,B_1,...d_S,B_S}$, it follows that
\begin{equation} \label{EQN: Conditional state}
\rho_{dB}^r = \frac{1}{p_{ACC}^r} Tr_{r^c}\left(\bigotimes_{i\neq r }M_{ACC_i} \rho_{d_1,B_1,...d_S,B_S}\right),
\end{equation}
where $Tr_{r^c}$ indicates trace over all systems but $r$. Putting this together we have,

\begin{align} \label{EQN: proof Interactive}
Tr \left(\Pi_{dB}^\perp \rho_{dB} \right) &= \frac{1}{S} \sum_{r=1}^S Tr \left(\Pi_{d_r,B_r}^\perp \otimes_{i\neq r} M_{ACC_i}\rho_{d_1,B_1,...d_S,B_S} \right) \nonumber \\
& \leq Tr \left(Q \rho_{d_1,B_1,...d_S,B_S}\right),
\end{align}
where $Q= \sum_{r=1}^S \Pi_{d_r,B_r}^\perp \otimes_{i\neq r} \frac{\left(I_{d_i}\otimes I_{B_i} + \Pi_{d_i,B_i}\right)}{2}$. It can easily be seen that $Q$ has maximum eigenvalues of $1/S$, reached by projection $\Pi_{d_j,B_j}^\perp \otimes_{i\neq j}  \Pi_{d_i,B_i}$ for any $j$. With this, we arrive at the following theorem.

\begin{theorem} \label{TH: Int}
The interactive protocol defined above is $\epsilon$-secure, with scaling $\epsilon = 1/S$.
\end{theorem}

Note here that the scaling of the protocol is inverse linear, as compared to exponential in \cite{BCGST02}. This can be understood as the cost of making the protocol practical. To get the exponential scaling in \cite{BCGST02} they require  entangled measurements or encodings over $S$ systems (which quickly becomes infeasible), whereas our protocol requires $S$ \emph{copies} of the single round encodings, which adds no difficulty in standard optical implementations, and is implementable with current technology.

\bigskip
The protocol above suffers from two main issues. Firstly, interaction is needed between the dealer and the players. Although this could be allowed in principle, it is more interesting if limited or no interaction is needed. Second, largely due to the interaction, memory is required by the dealer and players $B$ between steps \ref{PInt: IntAuthProc step1} and \ref{PInt: Teleportation}.  The dealer must keep their part of the EPR pairs until the players have recieved their shares, and $B$ must keep their shares until $r$ is announced by the dealer and further until $d$ announces their result $x$ for the teleportation. These problems can be overcome by replacing communication and the entanglement between the dealer and players with shared random keys, as was done in \cite{BCGST02}, but with an extra twist - they should be shared using a classical secret sharing protocol, so that the access structure is maintained. In this way, the protocol below requires no interaction after the initial sharing of a random key and the dealer's use of the channel, and similarly no memory is required by $B$ or $d$ and no entanglement is needed between $d$ and $P$ (it is of the `prepare and measure' type). To encode the randomly chosen $r$, we define string $q=(q_1,...q_S)$ such that $q_i=0$ if $i\neq r$, $q_i=1$ if $i=r$, where $r$ is randomly chosen in $[1,..S]$.


\bigskip

\noindent{\bf  Non-interactive Protocol}
\begin{enumerate}[topsep=2pt, partopsep=0pt,itemsep=0mm]
\item $d$ and $P$ share random strings $q,t,y,x$ via a classical secret sharing scheme over P (i.e. $d$ knows each string, but it is shared via a classical secret sharing scheme with the relevant access structure over $P$ so that only authorised sets can access it, and only when they collaborate to do so, and unauthorised sets get no information at all).

\item Going through round by round $i=1...S$. If $q_i=0$ the dealer proceeds to step \ref{PNonInt: test}, if $q_i=1$ the dealer proceeds to step \ref{PNonInt: use}. \label{PNonInt: Choose}

\item \label{PNonInt: test}
For $q_i = 0$
\begin{enumerate}[topsep=0pt, partopsep=0pt,itemsep=0mm]

\item Dealer prepares and distributes state $H_L^{t_i}Z_L^{y_i}\frac{|0_L\rangle_{P}+|1_L\rangle_{P}}{\sqrt{2}}$.

\item After receiving the state from the dealer, authorised set $B$ collaborate to find $q_i$ (which is $0$), $t_i$, and $y_i$.

\item Authorised set $B$ measures $X_{L,B}$ if $t_i=0$ or measures $Z_{L,B}$ if $t_i=1$ . The result is denoted $y_i'$. \\If $y_i\neq y_i'$ REJECT. If $y_i=y_i'$ ACCEPT.

\item If $i = S$, END, otherwise return to step \ref{PNonInt: Choose}.

\end{enumerate}

\item \label{PNonInt: use}
For $q_i=1$
\begin{enumerate}[topsep=0pt, partopsep=0pt,itemsep=0mm]
\item $d$ encodes and distrubutes the state $X_L^{x_0}Z_L^{x_1}|\psi_L\rangle_P$.

\item After recieving the state from the dealer, authorised set $B$ collaborate to find $q_i$ (which is $1$), $t_i$, $y_i$ and $x$.

\item $B$ decodes using $X_{L,B}, Z_{L,B}$.

\item If $i = S$, END, otherwise return to step \ref{PNonInt: Choose}

\end{enumerate}

\end{enumerate}

\bigskip
Replacing the communication by shared random strings in this way does not effect the security \cite{BCGST02,SP00}, and we have the following theorem.

\begin{theorem} \label{TH: NonInt}
The non- interactive protocol defined above is $\epsilon$-secure, with $\epsilon = 1/S$.
\end{theorem}

\section{Example}\label{QSS}

In recent years graph states have emerged as a useful framework in which to do secret sharing \cite{MS08,KFMS10,MMP13,MM13}.
As an example, we now illustrate how secret sharing over untrusted channels works for the case of five players, such that any set of three or more players can access the secret and any fewer have none (the so called $(3,5)$ threshold secret sharing scheme of \cite{MS08}). To begin, we introduce some notation. A graph state $|G\rangle_{1,...,n}$ is a state on $n$ qubits which is associated to graph $G$ through graph state stabiliser operators $K_i:=X_i\otimes_{j \in N(i)} Z_j$ where $i$ is associated to a vertex in the graph and $N(i)$ are the set of its neighbours, and the eigenequations $K_i|G\rangle_{1,...n}= |G\rangle_{1,...n}$ $\forall i$.

In our example the logical states are given by $|0_L\rangle_P=|G_P\rangle_P$ for the graph $G_P$ in Fig.~\ref{FIG: (3,5)}a), and $|1_L\rangle_P=Z_1Z_2Z_3Z_4Z_5 |G_P\rangle_P$ with $P=\{1,2,3,4,5\}$. The entangled state used in the interactive protocol step \ref{PInt: IntAuthProc step1} is
\begin{eqnarray}
|\Phi\rangle_{dP}=\frac{|0\rangle_d|0_L\rangle_P+|1\rangle_d|1_L\rangle_P}{\sqrt{2}}.
\end{eqnarray}
It is not difficult to see that this is itself a graph state associated to the graph in Fig.~\ref{FIG: (3,5)}b), $|\Phi\rangle_{dP}=|G_{dP}\rangle_{dP}$.

The choice of logical operators depends on the set $B\subset P$ who are trying to access the secret. Notice that during the protocol the dealers' action does not require knowledge of $B$ - this is essential to secret sharing, so that the players can decide for themselves who access the secret.

If players $B=\{1,2,3\}$ wish to act as the authorised set, they can use logical operators $X_L=X_1 Z_2 X_3$ and $Z_L=Z_1 X_2 Z_3$. The logical operators measured in the test step of the protocol (step \ref{PInt: Test} for the interactive and step \ref{PNonInt: test} for the non-interactive) are graph state stabilisers of the graph $G_{dB}$ given in Fig.~\ref{FIG: (3,5)}c), $X_d \otimes X_{L,B} = K_d K_1 K_3 $, $Z_d \otimes Z_{L,B} = K_2$. This is no coincidence, and is a general feature of graph state protocols, which gives a simple decomposition of states into the graph state basis as the natural expansion for (\ref{EQN: logical expansion}). To decode the secret after teleportation, players $1$ and $3$ measure in the Bell basis, and the secret is passed onto the qubit of player $2$. This is the same decoding procedure for the standard secret sharing protocol \cite{MS08}.

If players $B=\{1,3,4\}$ wish to act as the authorised set, they can use logical operators $X_L=Z_1 X_3 X_4$ and $Z_L=X_1 X_3 X_4$. The logical operators measured in the test step of the protocol (\ref{PInt: Test} for the interactive and \ref{PNonInt: test} for the non-interactive) are graph state stabilisers of the graph $G_{dB}$ given in Fig.~\ref{FIG: (3,5)}d), $X_d \otimes X_{L,B} = K_d K_3 K_4 $, $Z_d \otimes Z_{L,B} = K_1 K_3 K_4$. In this case, to decode the secret after teleportation, players $3$ and $4$ measure in the Bell basis, and the secret is passed onto the qubit of player $1$.

Logical operators can similarly be defined for all sets of three players, and it is easy to see that any two players cannot access the secret by themselves \cite{MS08}.

\begin{figure}[t!]
a) \begin{tikzpicture}[shorten >=1, -, font=\footnotesize,scale=0.8]
\tikzstyle{nonode}=[circle, draw, fill=black!10, inner sep=0pt, minimum width=10pt]
   \draw \foreach \x/\name in {0/2,72/1,144/5,216/4,288/3}
  {(\x+19:1) node[nonode] {\footnotesize{$\name$}}  -- (\x+91:1)};

\end{tikzpicture}\hspace{2cm}
b) \begin{tikzpicture}[shorten >=1, -, font=\footnotesize,scale=0.8]
    \tikzstyle{nonode}=[circle, draw, fill=black!10, inner sep=0pt, minimum width=10pt]
\node[nonode] (d) at (0,0)  {\footnotesize{$d$}};
\draw \foreach \x/\name in {0/2,72/1,144/5,216/4,288/3}
      {(d) --  (\x+19:1) node[nonode] {\footnotesize{$\name$}}  -- (\x+91:1)};
  \end{tikzpicture}\hspace{2cm}

c)\begin{tikzpicture}[shorten >=1, -, font=\footnotesize,scale=0.8]
\tikzstyle{nonode}=[circle, draw, fill=black!10, inner sep=0pt, minimum width=10pt]
\node[nonode] (d) at (0,0)  {\footnotesize{$d$}};
   \draw \foreach \x/\name in {0/2,288/3}
  {(d) -- (\x+19:1) node[nonode] {\footnotesize{$\name$}}  -- (\x+91:1)};
\draw (d) -- (91:1) node[nonode] {\footnotesize{$1$}};

\end{tikzpicture}\hspace{2cm}
d)\begin{tikzpicture}[shorten >=1, -, font=\footnotesize,scale=0.8]
    \tikzstyle{nonode}=[circle, draw, fill=black!10, inner sep=0pt, minimum width=10pt]
\node[nonode] (d) at (0,0)  {\footnotesize{$d$}};
\draw (235:1) -- (307:1);
\draw \foreach \x/\name in {72/1,216/4,288/3}
      {(d) --  (\x+19:1) node[nonode] {\footnotesize{$\name$}}};

\end{tikzpicture}
\caption{Graphs for sharing a secret amongst five players such that any majority can access the secret \cite{MS08}.a) Graph $G_P$ associated to logical encoding. b) Graph $G_{dP}$ for the entangled state between $d$ and $P$ used in step 1 of the protocol. c) Graph $G_{dB}$ associated to the test for players $B=\{1,2,3\}$. d) Graph $G_{dB}$ associated to the test for players $B=\{1,3,4\}$.}\label{FIG: (3,5)}
\end{figure}

\section{Conclusions}

Several variations of these protocols are also possible. In particular, if the quantum secret is precious, the dealer may not want to send the information when the channel is not trusted - i.e. when it fails the test part of the protocols above. To address this one can adapt the protocol to allow for the players to announce abort when they fail the test. It is possible to show an adapted theorem for security, which is only slightly modified. We present the protocol and the theorem in the appendix \ref{SCN: App Abort protocol}. One subtle issue with how this may be used however, is that the protocol is now interactive (albeit limited to when the players announce abort). When the accessing set $B$ announce abort they declare themselves, which may be an issue in some uses of secret sharing as a primitive (one may try to develop ways to overcome this for example using an anonymous announcement of abort).
Another alternative protocol with abort can be found if, instead of randomly choosing in a fixed number ($S$) of rounds, at each round we randomly choose to test or use the channel. This allows for slightly different security statements, as used in \cite{PCW12} for entanglement verification. We consider this simplified unbounded aborting protocol in appendix \ref{SCN: App Abort protocol}.

Using graph state schemes as in the example shown here has several advantages. Firstly they are a common framework for many quantum information processing tasks, including measurement based quantum computation and error correction. This allows these protocols to naturally fit into more general and elaborate network scenarios integrating several of these tasks. Furthermore, this relationship allows us to understand the connection between different protocols. The measurements used for the test part of our protocol are exactly those used for the $CQ$ protocols to establish secure key between the dealer and authorised players \cite{MM13}. This relationship is general - whenever complementary bases are used to establish a secure key, successful key generation implies the channel works \cite{CW05,MM13}.

Secondly graph states are very well suited to implementation. Many schemes exist for the generation and manipulation of graph states in different technologies including linear optics \cite{Browne05}, continuous variables \cite{Rigas12,Ukai11} and ion traps \cite{Wunderlich09,Lanyon13}. Optical implementations of graph states are ideal for the secret sharing protocols presented here, and experiments in this direction are well advanced, indeed optics has been used to implement sophisticated quantum information processing tasks including measurement based quantum computing \cite{WRR05}, blind quantum computation \cite{BKBFW12}, and error correction \cite{BHT14}. These experiments contain all the key steps needed for our protocols.

All of the results here easily extend to the qudit case, including use for qudit graph state secret sharing \cite{KFMS10,MM13}, which allows our scheme to be used to cover all access structures. One simply extends the states and operators to their natural qudit versions. This is done explicitly in the appendix \ref{SCN: Qudit}. Furthermore, the proofs also follow through for secret sharing protocols using mixed state encodings, as well as the hybrid protocols of \cite{BCT09},\cite{JMP11} where classical secret sharing is used in addition to allow for access structures otherwise forbidden, in this case however authenticated classical channels between the dealer and the players must also be secure.

\bigskip

\noindent {\bf Acknowledgements.} We thank Anthony Leverrier, Eleni Diamanti and Andr\'{e} Chailloux for many useful discussions and  helpful comments. This work was supported by the Ville de Paris Emergences program, project CiQWii.

\appendix

\section{Aborting protocols} \label{SCN: App Abort protocol}
We first present an adaptation of the non-interactive protocol to include abort.\\
\\
\noindent{\bf Protocol with Abort}
\begin{enumerate}[topsep=2pt, partopsep=0pt,itemsep=0mm]
\item $d$ and $P$ share random strings $q,t,y,x$ via a classical secret sharing scheme over P.

\item Going through round by round $i=1...S$. Authorised set $B$ collaborate to find $q_i$. If $q_i=0$ all proceed to step \ref{P3: test}, if $q_i=1$ all proceed to step \ref{P3: use}. \label{P3: Choose}

\item \label{P3: test}
For $q_i = 0$
\begin{enumerate}[topsep=0pt, partopsep=0pt,itemsep=0mm]

\item Dealer prepares and distributes state $H_L^{t_i}Z_L^{y_i}\frac{|0_L\rangle_{P}+ |1_L\rangle_{P}}{\sqrt{2}}$.

\item For authorised set $B$, if $t_i=0$, $B$ measures $Z_L$, if $t_i=1$, $B$ measures $X_L$. The result is denoted $y_i'$. \\
If $y_i\neq y_i'$ $B$ announces ABORT, all abort.

\item If $y_i=y_i'$ ACCEPT, and return to step \ref{P3: Choose}.

\end{enumerate}

\item \label{P3: use}
For $q_i=1$
\begin{enumerate}[topsep=0pt, partopsep=0pt,itemsep=0mm]
\item $d$ encodes and distributes the state $X_L^{x_0}Z_L^{x_1}|\psi_L\rangle_P$.

\item $B$ decode.

\item END

\end{enumerate}

\end{enumerate}

\begin{theorem}
The aborting protocol defined above is $\epsilon$-secure, with $\epsilon < 2/S$.
\end{theorem}

The proof follows the same argumentation as before, but in this case the testing stops at round $r$. This gives a different $Q$ operator (applied as in equation (\ref{EQN: proof Interactive})),
\begin{equation}
Q=\frac{1}{S}\sum_{r=1}^{S} \Pi_{d_r,B_r}^\perp \bigotimes_{i < r} \frac{\left(I_{d_i}\otimes I_{B_i} + \Pi_{d_i,B_i}\right)}{2} \bigotimes_{i>r} (I_{d_i}\otimes I_{B_i}).
\end{equation}
To find the greatest eigenvalue we can decompose into the $\{\Pi_{d_i,B_i}, \Pi_{d_i,B_i}^\perp\}$ basis. Then we have a sum with all strings of products, where each string $x$, in which $\Pi_{d_i,B_i}$ appears $|x|$ times, occurs with weight $1/S \sum_{a=0}^{|x|-1}1/2^a$. Thus, the greatest eigenvalue of $Q$ is $1/S\sum_{a=0}^{S-1}1/2^a = \frac{2-2^{-S}}{S} <2/S$, given by the string $\Pi_{d_1,B_1}^\perp \otimes \Pi_{d_2,B_2}^\perp \otimes ... \Pi_{d_S,B_S}^\perp$. In this case we see that the optimal cheat (in terms of maximising the failure of the protocol) is given by the dishonest parties preparing many copies of states all outside the ideal space.

\bigskip
We also present an aborting protocol where the randomness (over whether to test the channel or use it) is inserted each round.\\
\\
{\bf  Aborting  protocol, indefinite length}

\begin{enumerate}[topsep=2pt, partopsep=0pt,itemsep=0mm]
\item \label{IntAuthProc with abort step1}
Dealer $d$ generates EPR state $|\Phi\rangle_{dP}$ and sends the shares to $P$.

\item After $P$ received all their parts, $d$ chooses $r\in [0,...S-1]$ at random and sends $r$ to $P$. If $r\neq 0$, continue to step \ref{IntAuthProc with abort test}, otherwise move to step \ref{IntAuthProc with abort use}.

\item \label{IntAuthProc with abort test}
TEST
\begin{enumerate}[topsep=0pt, partopsep=0pt,itemsep=0mm]
\item $d$ chooses $t\in\{0,1\}$. If $t=0$, $d$ measures $Z_d$, if $t=1$ $d$ measures $X_d$. The result is denoted $y$. $d$ sends $t$ and $y$ to $P$.

\item If $t=0$, $B$ measures $Z_{L,B}$, if $t=1$, $B$ measures $X_{L,B}$. The result is denoted $y'$.
If $y_i\neq y_i'$ $B$ announces ABORT, all abort.

\item If $y_i=y_i'$ ACCEPT, and return to step \ref{IntAuthProc with abort step1}.
\end{enumerate}

\item \label{IntAuthProc with abort use}
USE CHANNEL
\begin{enumerate}[topsep=0pt, partopsep=0pt,itemsep=0mm]
\item $d$ uses EPR pair to teleport the secret state onto the logical basis, denoting the bell basis measurements $x$, and sends $x$ to all $P$. Upon receiving $x$, $B$ decodes.
\end{enumerate}
\end{enumerate}

\bigskip

This protocol is similar to the use of the verification of GHZ states in \cite{PCW12}, and we can present a similar security statement as there. Let $C_f$ be the event that the state is teleported, and that the fidelity of any teleported state $\rho$ and the sent state $|\psi\rangle$ is bounded by $\langle \psi |\rho |\psi\rangle \leq f$, then we have the following theorem.

\begin{theorem}
For all f, the probability of event $C_f$ is bounded
 \begin{equation}
 P(C_f)) \leq \frac{2}{S(1-f)}.
 \end{equation}
\end{theorem}

To prove this, call $C_f^N$ the event that at round $N+1$ the state is used (which means that all previous rounds have returned accept) and that the fidelity of all states sent satisfies $\langle \psi |\rho |\psi\rangle \leq f$. The probability of this event is given by the probability of using the $N+1$th round state, times the probability of having tested the previous rounds times the probability of passing all the previous rounds.
\begin{equation}
P(C_f^N) = \frac{1}{S} \left( 1-\frac{1}{S}\right)^N \prod_{i=1}^N p_i
\end{equation}
where $p_i$ is the probability that round $i$ is accepted. This is given by
\begin{align}
p_i &= Tr(\rho_{d_i,B_i} M_{ACC_i})  \nonumber\\
& \leq \frac{1+ Tr(\rho_{d_i,B_i} \Pi_{d_i,B_i})}{2}\nonumber \\
& \leq \frac{1+ f}{2}.
\end{align}

The probability of event $C_f$ is then given by taking the limit of the sum over $N$, which is bounded by taking the integral.
\begin{align}
P(C_F) &\leq \frac{1}{S} \sum_{0}^{\infty}\left( 1-\frac{1}{S}\right)^N \left(\frac{1+f}{2}\right)^N \nonumber \\
&\leq \frac{1}{S} \int_{N=0}^\infty \left(1- \left(\frac{1-f}{2}\right)\right)^N dN \nonumber \\
&= \frac{1}{S}\frac{-1}{\log\left(1- \left(\frac{1-f}{2}\right)\right)}\nonumber\\
 &\leq \frac{2}{S(1-f)}.
\end{align}

\section{Qudit protocols} \label{SCN: Qudit}

The qudit versions work in the same way, by simply replacing states and operators by their generalised high dimensional versions. For simplicity we consider prime dimension $q$ (this is sufficient for allowing for all access structures \cite{BCGST02}). Paulis and the computational basis are replaced by their qudit extensions, $X$, $Z$, $X|i\rangle = |i \oplus 1\rangle$ (where now $\oplus$ denotes sum modulo $q$) and $Z|i\rangle=\omega^i |i\rangle$, $\omega=e^{i2\pi/q}$ (analogously for all the possible sets of logical operators and basis states), and the EPR state by its qudit version
\begin{equation}
|\Phi\rangle_{dP} = \frac{1}{\sqrt{q}}\sum_{i=0}^{q-1}|i\rangle_d |i_L\rangle_P.
\end{equation}

We similarly define the projection operator on $d$ and subset of players $B\subset P$,
\begin{equation}
\Pi_{dB} = \sum_{t\in F_q^2} Z_d^{t^1}X_d^{t^2} \otimes Z_{L,B}^{t^1}X_ {L,B} ^{t^2},
\end{equation}
where $t=\{t^1,t^2\}$, $t^i\in F_q$. This again defines a space of maximally entangled states between the dealer and players $B$.

For simplicity we present a qudit protocol which includes all the measurements in this projection. This is not necessary for our results, indeed for the qubit case we had fewer and similar results follow using fewer measurements as in the qubit version - it is mostly a matter of taste if one chooses the full set or fewer (similar to the situation for qudit versions of QKD \cite{Sheridan10}). We adopt it here for its simpler presentation. It allows for analagous statement of Theorems \ref{TH: Int}, \ref{TH: NonInt} following the same logic as for the qubit version.

\bigskip

\noindent{\bf Interactive Protocol (qudit)}
\begin{enumerate}[topsep=2pt, partopsep=0pt,itemsep=0mm]
\item Dealer $d$ generates $S$ qudit EPR states, $|\Phi\rangle_{dP}^{\otimes S}$, and sends the shares of each one to $P$. \label{PInt: IntAuthProc step1 qudit}

\item After $P$ received all their parts, $d$ chooses $r\in [1,...S]$ at random and sends $r$ to $P$.

\item For qudit EPR pairs  $i\neq r$, $d$ chooses $ t_i\in F_q^2 $, and measures $Z_{d}^{t^1_i}X_{d}^{t^2_i}$ , and denotes the result $y_i$, and sends $t_i$ and $y_i$ to $P$.

\item For qudit  EPR pairs $i\neq r$, accessing set $B$ measure $Z_{L,B}^{t^1_i}X_{L,B} ^{t^2_i}$ . Denoting result by $y_i'$ if $y_i=y_i'$ ACCEPT, if $y_i\neq y_i'$ REJECT.\label{PInt: Test qudit}

\item For $i=r$, $d$ uses qudit EPR pair $r$ to teleport the secret state onto the logical basis, denoting the qudit bell basis measurements $x$, and sends $x$ to all $P$. Upon receiving $x$, $B$ decodes.\label{PInt: Teleportation qudit}
\end{enumerate}

\begin{theorem} \label{TH: prob pass qudit}
Any state $\rho_{dB}$ with fidelity $F=Tr(\Pi_{dB}\rho_{dB})$ to projector $\Pi_{dB}$ accepts at step $\ref{PInt: Test}$ of the qudit interactive protocol with probability $\D Pr = \frac{1+F}{2}$ .
\end{theorem}

\begin{theorem}\label{TH: Int Protocol qudit}
The interactive qudit protocol defined above is $\epsilon$-secure, with $\epsilon = 1/S$.
\end{theorem}

The non-interactive version follows similarly, as does the Theorem. The state $|\Psi^{t,y}\rangle_{P}$ denotes the eigenstate of the operator $ Z_{L}^{t^1_i}X_{L}^{t^2_i}$ with eigenvalue $\omega^y$.

\bigskip

\noindent{\bf  Non-interactive Protocol (qudit)}
\begin{enumerate}[topsep=2pt, partopsep=0pt,itemsep=0mm]
\item $d$ and $P$ share random strings $r,t,y,x$ via a classical secret sharing scheme over P (i.e. $d$ knows the string, but it is shared via a classical secret sharing scheme with the relevant access structure over $P$ so that only authorised sets can access it, and only when they collaborate to do so, and unauthorised sets get no information at all).

\item Going through round by round $i=1...S$. If $q_i=0$ the dealer proceeds to step \ref{PNonInt: test qudit}, if $q_i=1$ the dealer proceeds to step \ref{PNonInt: use qudit}. \label{PNonInt: Choose qudit}

\item \label{PNonInt: test qudit}
For $q_i = 0$
\begin{enumerate}[topsep=0pt, partopsep=0pt,itemsep=0mm]

\item Dealer prepares and distributes state $|\Psi^{t,y}\rangle_{P}$.

\item After receiving the state from the dealer, authorised set $B$ collaborate to find $q_i$ (which is $0$), $t_i$ and $y_i$.

\item Authorised set $B$ measures $Z_{L,B}^{t^1_i}X_{L,B}^{t^2_i}$ . The result is denoted $y_i'$. \\If $y_i\neq y_i'$ REJECT. If $y_i=y_i'$ ACCEPT.

\item If $i = S$, END, otherwise return to step \ref{PNonInt: Choose qudit}.

\end{enumerate}

\item \label{PNonInt: use qudit}
For $q_i=1$
\begin{enumerate}[topsep=0pt, partopsep=0pt,itemsep=0mm]
\item $d$ encodes and distrubutes the state $X_L^{x_0}Z_L^{x_1}|\psi_L\rangle_P$.

\item After recieving the state from the dealer, authorised set $B$ collaborate to find $q_i$, which is $1$.

\item $B$ decodes.

\item If $i = S$, END, otherwise return to step \ref{PNonInt: Choose qudit}

\end{enumerate}

\end{enumerate}

\bigskip

\begin{theorem} \label{TH: PNonInt qudit}
The non-interactive qudit protocol defined above is $\epsilon$-secure, with $\epsilon = 1/S$.
\end{theorem}

The aborting protocol follows similarly.

\end{document}